\documentclass[twocolumn,preprintnumbers,amsmath,amssymb]{revtex4}
\usepackage{graphicx}
\usepackage{dcolumn}
\usepackage{bm}
\usepackage{color}
\usepackage[final]{pdfpages}
\begin{document}
\title{Thermal transport and phonon hydrodynamics in strontium titanate}
\author{Valentina Martelli$^{1}$, Julio Larrea Jim\'{e}nez$^{2}$,  Mucio Continentino$^{1}$, Elisa Baggio-Saitovitch$^{1}$  and Kamran Behnia$^{3,4}$}
\affiliation{(1)Centro Brasileiro de Pesquisas F\'{\i}sicas, 22290-180, Rio de Janeiro, RJ, Brazil\\
(2)Institute of Physics, University of S\~{a}o Paulo, CEP 05508-090, S\~{a}o Paulo, SP, Brazil\\
(3) Laboratoire Physique et Etude de Mat\'{e}riaux (CNRS-UPMC), ESPCI Paris, PSL Research University, 75005 Paris, France\\
(4)II. Physikalisches Institut, Universit\"{a}t zu K\"{o}ln,  50937 K\"{o}ln, Germany}
\date{January 26, 2018}

\begin{abstract}
We present a study of thermal conductivity, $\kappa$, in undoped and doped strontium titanate in a wide temperature range (2-400 K) and detecting different regimes of heat flow. In undoped SrTiO$_{3}$, $\kappa$ evolves faster than cubic with temperature below its peak and in a narrow temperature window. Such a behavior, previously observed in a handful of solids, has been attributed to a Poiseuille flow of phonons, expected to arise when momentum-conserving scattering events outweigh momentum-degrading ones. The effect disappears in presence of dopants. In SrTi$_{1-x}$Nb$_{x}$O$_{3}$, a significant reduction in lattice thermal conductivity starts below the temperature at which the average interdopant distance and  the thermal wavelength of acoustic phonons become comparable. In the high-temperature regime, thermal diffusivity becomes proportional to the inverse of temperature, with a prefactor set by sound velocity and Planckian time ($\tau_{p}=\frac{\hbar}{k_{B}T}$).
\end{abstract}
\maketitle

Heat travels in insulators thanks to phonons. This has been described by the Peierls-Boltzmann equation, which quantifies the spatial variation in phonon population caused by the temperature gradient. In recent years, thanks to improved computing performance and new theoretical techniques, a quantitative account of intrinsic thermal conductivity of semiconductors is accessible to first-principle theory\cite{Lindsay:2013}. When most scattering events conserve momentum and do not decay heat flux, collective phonon excitations, dubbed relaxons, become fundamental heat carriers \cite{Cepellotti:2016}. This hydrodynamic regime of phonon flow, identified decades ago\cite{Sussmann:1963,Guyer:1966,Gurzhi:1968,Beck:1974}, is gaining renewed attention in the context of graphene-like two-dimensional systems\cite{Cepellotti:2015,Lee:2015}.

\begin{figure}
\includegraphics[width=0.42\textwidth]{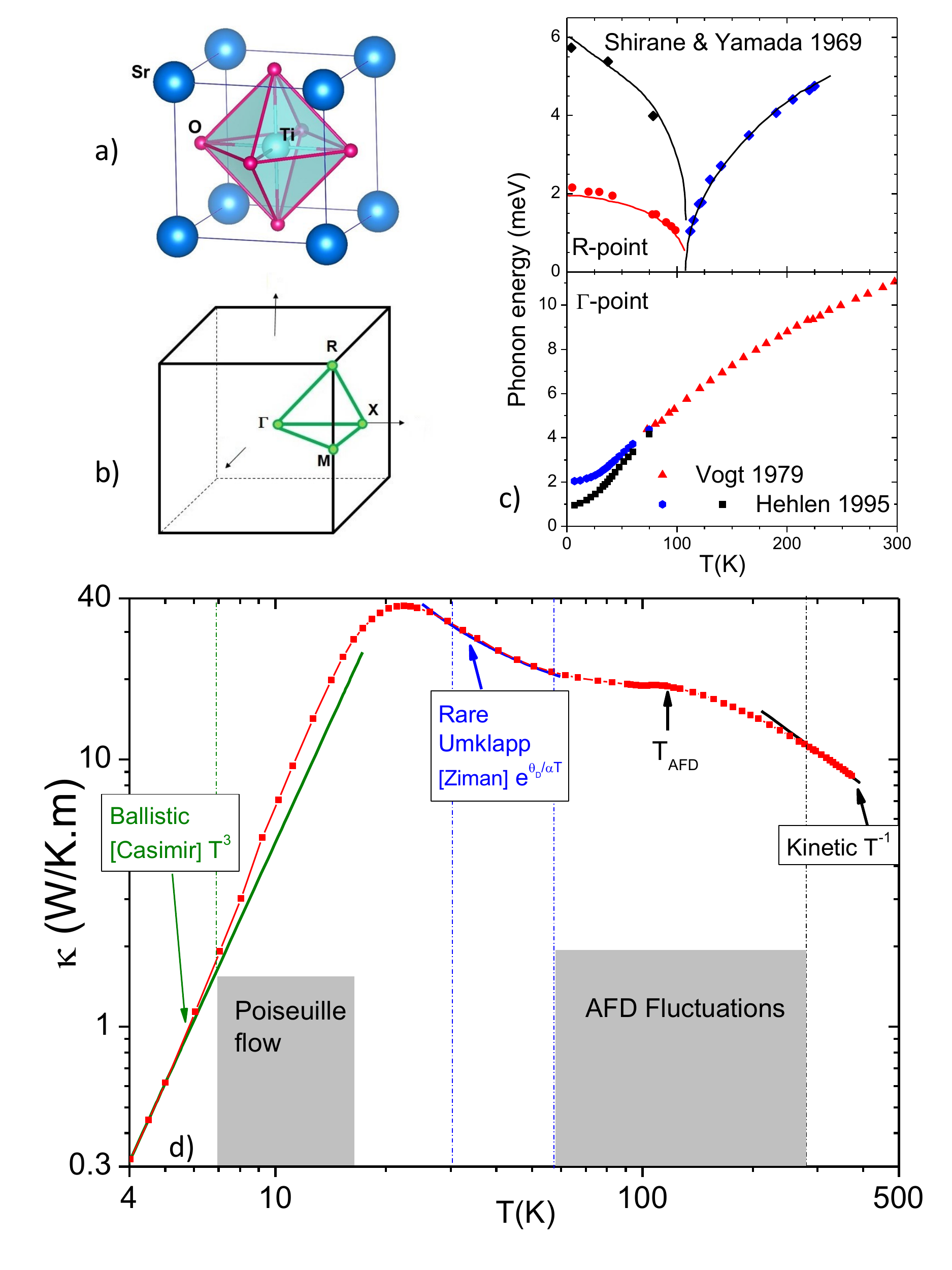}
\caption{ a) Crystal structure of strontium titanate; b) The cubic Brillouin zone and its high-symmetry points. c) The temperature dependence of the two soft modes according to the neutron scattering studies\cite{Shirane:1969}, hyper-Raman \cite{Vogt:1979} and Brillouin scattering spectroscopy\cite{Hehlen:1999}.  d) Thermal conductivity of a SrTiO$_{3}$ crystal(closed red squares) in a log-log plot (For a linear plot see Fig. 3a). Different regimes of thermal transport  are identified.  Solid lines represent the expected behaviors in these regimes.  An additional window due to enhanced Umklapp scattering opens up in the vicinity of the Antiferrodistortive (AFD) transition.}
\label{Fig1}
\end{figure}

The perovskyte SrTiO$_{3}$ is a quantum paraelectric\cite{Muller:1979}, which owes its very existence to zero-point quantum fluctuations. First-principle calculations find imaginary phonon modes\cite{Aschauer:2014}, which hinder a quantitative understanding of the lattice thermal transport \cite{Steigmeier:1968}. This insulator turns to a metal upon the introduction of a tiny concentration of dopants. The metal has a dilute superconducting ground state\cite{Lin:2013} and an intriguing room-temperature charge transport\cite{Lin:2017}. Its thermal conductivity has remained largely unexplored, in contrast to electric\cite{Lin:2015} and thermoelectric\cite{Cain:2013} transport.

In this Letter, we present an extensive study of thermal conductivity, $\kappa$, of undoped and doped SrTiO$_{3}$ crystals and report on three new  findings. First of all, in a narrow temperature range, thermal conductivity evolves faster than cubic. This behavior had only been reported in a handful of solids\cite{Beck:1974} and attributed to a Poiseuille flow of phonons. We argue that the emergence of phonon hydrodynamics results from the multiplication of  momentum-conserving scattering events  due to the presence of a ferroelectric soft mode, as suggested decades ago \cite{Gurevich:1988}. It lends support to previous reports on the observation of the second sound in this system\cite{Hehlen:1995,Koreeda:2007}, which has been controversial\cite{Scott:2000}. Second, our study finds that a random distribution of dopants drastically reduces thermal conductivity below a temperature which tunes the heat-carrying phonon wavelength to the average interdopant distance. Finally, we put under scrutiny the thermal diffusivity of the system near room temperature  and link its magnitude  and temperature dependence to the so-called Planckian scattering time\cite{Bruin:2013}, in the context of the ongoing debate on a possible boundary to diffusivity\cite{Hartnoll:2015,Zhang:2017}.

The cubic elementary cell of strontium titanate encloses a TiO$_{6}$ octahedra and has strontium atoms at its vertices (Fig. 1a). Neutron and Raman scattering studies have identified two distinct soft modes. The first is associated with the antiferrodistortive (AFD) transition, which leads to the loss of cubic symmetry at 105 K\cite{Shirane:1969} by tilting two adjacent TiO$_{6}$ octahedra in opposite orientations. It is centered at the R-point of the Brillouin zone (Fig. 1b). The second soft mode \cite{Yamada:1969}, located at the zone center, is associated with the aborted ferroelectricity. Fig. 1c presents the temperature dependence of the two modes established by converging spectroscopic tools \cite{Shirane:1969,Vogt:1979,Hehlen:1999}. In common solids, only acoustic branches can host thermally-excited phonons at low temperatures. Here, phonons associated with these soft modes remain relevant down to fairly low temperatures.

We used a standard one-heater-two-thermometers technique to measure the thermal conductivity of  commercial single crystal of Sr$_{1-x}$Nb$_{x}$TiO$_{3}$\cite{supplement}. The results, presented in Fig.1d, reveal different regimes of heat transport classified by previous authors\cite{Guyer:1966,Beck:1974,Lee:2015}. Simply put, thermal conductivity is the product of specific heat, mean-free-path, and  velocity\cite{Berman:1976}. At one extreme, i.e. at low temperature, the phonon mean-free-path saturates, the system enters the ballistic  regime and $\kappa$ becomes cubic in temperature. In the other extreme,  at high temperature, the specific heat saturates and thermal conductivity, reflecting the temperature dependence of the mean-free-path, follows $T^{-1}$. In this kinetic regime, the wave-vector of thermally-excited phonons is large enough to allow Umklapp scattering events. Well below the Debye temperature, such events become rare and $\kappa$ increases exponentially. This is this Ziman regime.

The AFD transition has visible consequences for heat transport. First of all, it attenuates $\kappa$ near T$_{AFD}$, impeding a smooth evolution between T$^{-1}$ and exponential regimes. The R-point soft mode associated with the AFD transition provides additional Umklapp scattering at low energy cost. Interestingly, fitting $\kappa \propto exp(\frac{E_{D}}{T})$ in the Ziman regime, one finds $E_{D}\simeq$ 20 K, an energy scale comparable to the AFD soft mode. The second consequence of the AFD transition is to generate multiple tetragonal domains in an unstrained crystal\cite{Tao:2016}. Given that the typical size of tetragonal domains is a few microns\cite{Buckley:1999}, the upper boundary to the ballistic mean-free-path of phonons can be much lower than the sample dimensions.

\begin{figure}
\includegraphics[width=0.45\textwidth]{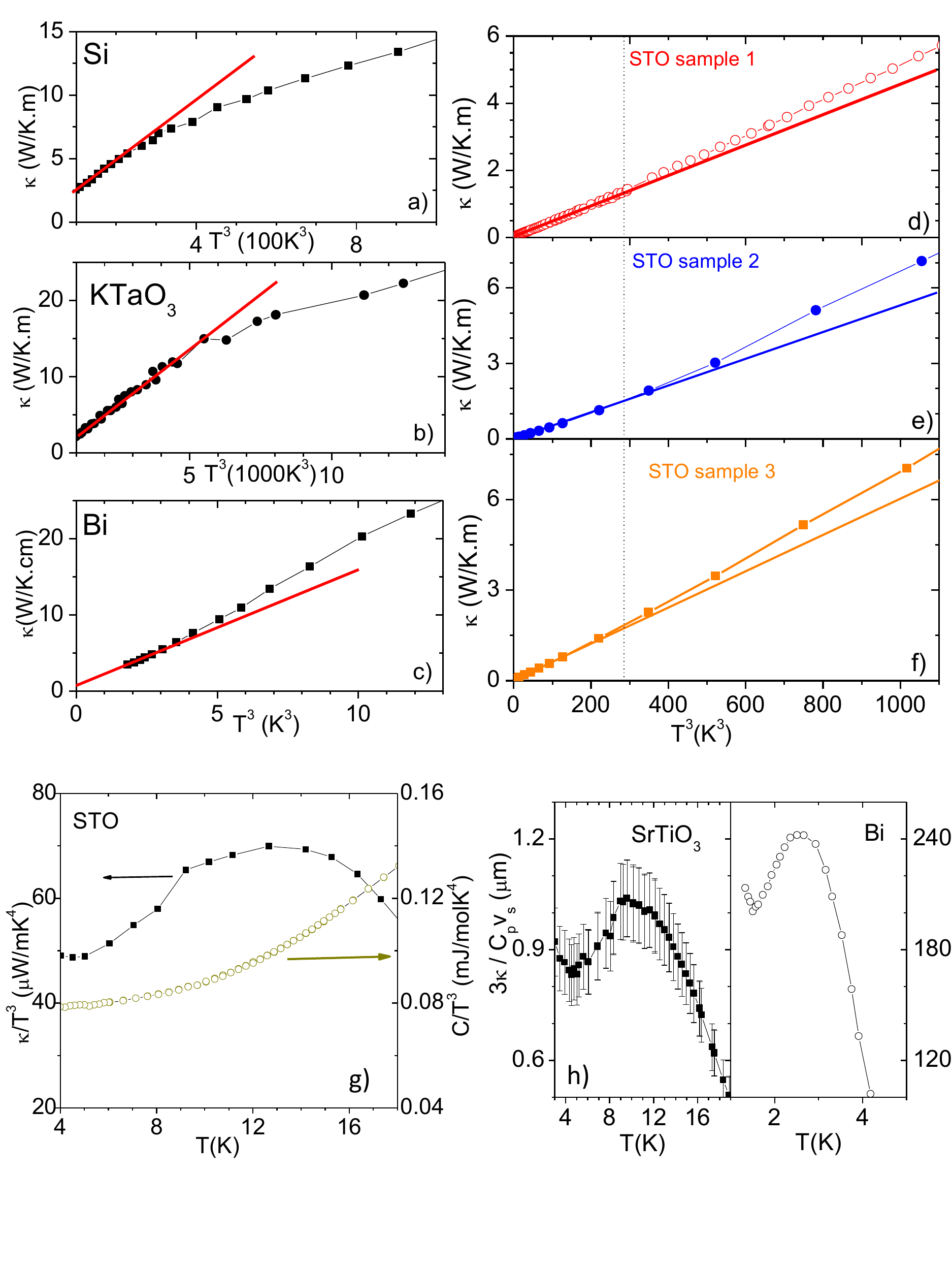}
\caption{Thermal conductivity, $\kappa$, as a function of $T^{3}$ in silicon (a) (after ref.\cite{Glass:1964}) and in KTaO$_{3}$(b).  In both, $\kappa$ deviates downward from the $T^{3}$ line. (c) In  bismuth (after ref. \cite{Kopylov:1974}) it deviates upward. In three different crystals of SrTiO$_{3}$ (d,e,f) the deviation is upward. g) Thermal conductivity and specific heat of SrTiO$_{3}$ evolve faster than cubic in this temperature range. But in a narrow window, thermal conductivity increases more rapidly. h) The apparent mean-free-path in both Bi and SrTiO$_{3}$ present a local peak, the hallmark of Poiseuille flow.}
\label{Fig2}
\end{figure}

We found a $\kappa$ varying faster than $T^{3}$ in a narrow (6 K $< T <$13 K) temperature window just below the peak. Usually, the ballistic regime ends with a downward deviation of $\kappa$ from its cubic temperature dependence. This happens in silicon\cite{Glass:1964} (Fig. 2a) or in KTaO$_{3}$ (Fig. 2b). This is not the case of bismuth where it shows an upward deviation between the ballistic regime and the peak (Fig. 2c). This has been identified as a signature of Poiseuille flow of phonons\cite{Kopylov:1974}.

The Poiseuille regime emerges when energy exchange between phonons is frequent enough to keep the local temperature well-defined and  Umklapp collisions are so rare that the flow is mainly impeded by boundary scattering. Without viscosity, no  external temperature gradient would be then  required to sustain the phonon drift\cite{Gurzhi:1968}. This picture, developed decades ago\cite{Sussmann:1963,Guyer:1966,Gurzhi:1968}, requires a hierarchy of time scales. The time separating two normal scattering events, $\tau_{N}$, should become much shorter than the time between boundary scattering events, $\tau_{B}$, and the latter much shorter than the time between resistive scattering events, $\tau_{R}$, which are due to either Umklapp or impurity scattering.  The same hierarchy ($\tau_{N}\ll\tau_{B}\ll\tau_{R}$) is required for second sound, a wave-like propagation of temperature and entropy, which has been observed in bismuth as well as in other solids displaying Poiseuille flow\cite{Beck:1974}.

We confirmed a faster than cubic $\kappa$ in three different SrTiO$_{3}$ crystals (Fig.2 d-f). Here, the identification of this behavior with Poiseuille flow  is less straightforward since the specific heat of SrTiO$_{3}$ also evolves faster than cubic between 4K and 20K\cite{Ahrens:2007}. This is because the Debye approximation is inadequate in the presence of soft modes and one needs to consider Einstein terms of the soft optical modes. In order to address this concern, we measured the specific heat of  our cleanest crystal and found that the thermal conductivity increases faster than specific heat (Fig. 2g). The effective mean-free-path, $\ell_{Ph}=\frac{3\kappa C_{p}}{v_{s}}$, extracted from the specific heat, $C_{p}$, and  the sound velocity, v$_s$, was found to show a peak comparable to what was found in bismuth\cite{Kopylov:1974} (Fig. 2h).  In both cases, $\ell_{Ph}$ presents a local maximum 1.3 times the Knudsen minimum. The magnitude of the latter is slighly smaller than the crystal dimensions in bismuth, and to the typical size of tetragonal domains in strontium titanate, which have been found to be of the order of a micrometer\cite{Buckley:1999}. As far as we know, the only available explanation for a local peak in $\ell_{Ph}$ is Poiseuille flow.

Neither in bismuth nor in strontium titanate, the chemical purity is exceptionally high. The same is true of black phosphorus, where a faster-than-cubic $\kappa$ was recently observed\cite{bp}. Therefore, in these cases, in contrast to He crystals, the Poiseuille flow is presumably caused by a large three-phonon phase space\cite{Lindsay:2008} for momentum-conserving (compared to momentum-degrading) scattering events. We note that the low-temperature validity of the $\tau_{N}\ll\tau_{R}$ inequality in strontium titanate was previously confirmed by low-frequency light-scattering experiments\cite{Koreeda:2007}. Anomalies  detected by Brillouin scattering experiments\cite{Hehlen:1999} are believed to be caused by strong anharmonic coupling between acoustic and optical modes at low temperatures. A strong hybridization between acoustic and transverse optical phonons was theoretically confirmed\cite{Bussmann:1997} and is expected to flatten the phonon dispersion.  This would pave the way for  frequent normal momentum exchange. It would also pull down the phonon velocity, providing an alternative explanation for an unusually short apparent mean-free-path.

Let us turn our attention to the effect of atomic substitution. Fig. 3a shows thermal conductivity of SrTi$_{1-x}$Nb$_{x}$O$_{3}$. The magnitude of $\kappa$ smoothly decreases with increasing dopant concentration. Only at lower temperatures, additional contribution by electrons outweighs the reduction in lattice thermal conductivity. In this range, we resolve a finite $T$-linear component in thermal conductivity of metallic samples due to the electronic component of thermal conductivity, $\kappa_{e}$. This is in agreement with a previous study focused on temperatures below 0.5 K\cite{Lin:2014}, which verified the validity of the Wiedemann-Franz(WF) law in the zero-temperature limit, namely: $\kappa_{e}\rho/T=L_{0}$, where $\rho$ is the electric resistivity and L$_{0}=2.45\times 10^{-8} V^{2}/K^{2}$ is the Lorenz number. Assuming the validity of the WF law at finite temperatures, one can separate the electronic, $\kappa_{e}$, and the phononic, $\kappa_{ph}$, components of the total thermal conductivity. At finite temperature, because of inelastic scattering, one expects $\kappa_{e}\rho/TL_{0}\leq 1$ and electric resistivity provides only a rough measure of $\kappa_{e}$, which, as seen in Fig. 3b,  becomes rapidly much smaller than $\kappa_{ph}$ with rising temperature.

\begin{figure}
\includegraphics[width=0.45\textwidth]{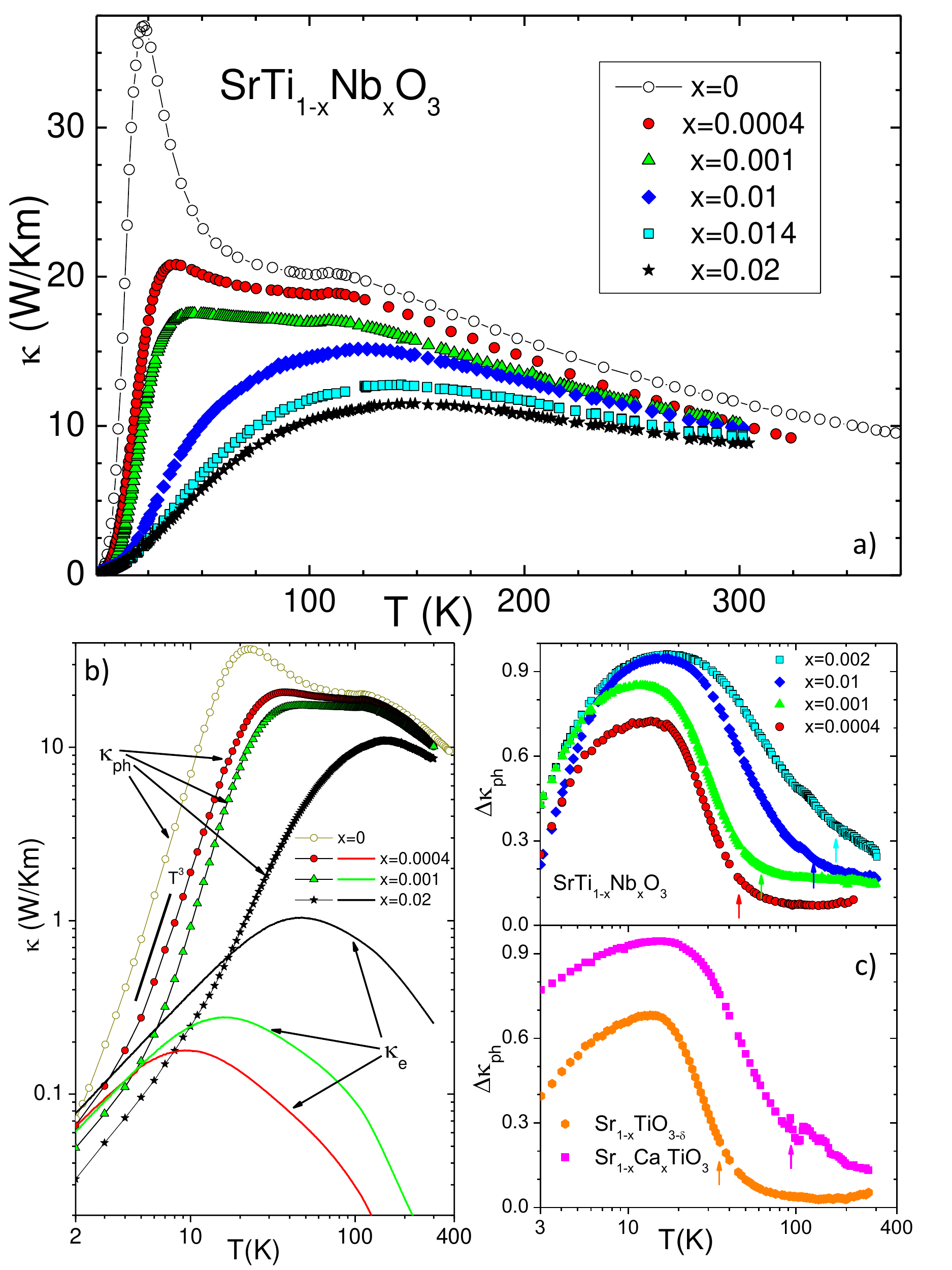}
\caption{ a) $\kappa$ as a function of temperature in  SrTi$_{1-x}$Nb$_{x}$O$_3$. b) Electronic, $\kappa_{e}$ and phononic, $\kappa_{ph}$, components of the thermal conductivity in three doped samples compared to undoped strontium titanate. Note the persistence of a $T^{3}$ behavior over a wide temperature window with a drastically reduced magnitude. c) Relative attenuation in phonon thermal conductivity, $\Delta\kappa_{ph}=1-\frac{\kappa_{ph} (x\neq0) }{\kappa_{ph} (x=0)}$ in SrTi$_{1-x}$Nb$_{x}$O$_3$  (top) and in Sr$_{1-x}$Ca$_{x}$ TiO$_3$ (x=0.0045) and in SrTiO$_{3-\delta}$ (n=7$\times$ 10$^{17}cm^{-3}$). Small arrows represent T$_{qn}$ (See text).}
\label{fig3}
\end{figure}

The first consequence of the disorder, introduced by this tiny substitution for $\kappa_{ph}$, is the loss of the faster than cubic regime associated with Poiseuille flow. As seen in Fig. 3b, reminiscent of what was observed in doped silicon and germanium\cite{Carruthers: 1957}, doping drastically damps $\kappa_{ph}$ at low temperature. The temperature-dependence of attenuation of phonon thermal conductivity caused by substitution: $\Delta\kappa_{ph}= 1- \kappa_{ph}(x\neq0)/ \kappa_{ph}(x=0)$, presented in  Fig. 3c displays a regular pattern. For small substitution (x= 0.0004), the  lattice thermal conductivity is reduced by 8 percent at room temperature, by as much as 70 percent at 20 K and by 20 percent at 3 K.  In other words, the maximum attenuation occurs in an intermediate temperature window. With increasing Nb concentration, the pattern is similar, but it shifts to higher temperatures. As seen in the lower panel of Fig. 3c, our measurements on an oxygen-reduced and a calcium-substituted sample produce similar patterns. Since Ca substitution\cite{Rischau:2017} keeps the system an insulator, one can conclude that the drastic reduction in lattice conductivity is mainly due to the random distribution of substituting atoms and \emph{not} to the scattering by mobile electrons.

A rigorous account of the temperature dependence of $\Delta\kappa_{ph}$ is missing.  We note, however, that $\Delta\kappa_{ph}$ drastically enhances at a temperature, which shifts upward as the the concentration increases (see upward arrows in Fig. 3c). Consider that with decreasing temperature, the typical wave-vector of thermally-excited phonons shrinks, following: $q_{ph}= \frac{k_{B}T}{\hbar v_{s}}$.   Therefore, at high-temperature, the phonon wave-length is shorter the average distance between dopants and the effect of disorder is limited.  The random distribution of dopants begins to matter  when the phonon wavelength becomes comparable to the average interdopant distance. In contrast to electrons, Anderson localization of phonons\cite{Luckyanova:2016} is not expected to impede diffusive transport\cite{Sheng:1994}. Theoretically, tiny level of disorder is sufficient to transform some  phonon modes from propagating waves (propagons) to diffusons, which travel diffusively, or to fully localized locons\cite{Seyf:2016}. One expects phonons with a wavelength much shorter or much longer than randomness length to be less affected. As a consequence, attenuation is to be more pronounced in the temperature window  where the most-concerned phonons happen to be dominant thermally-excited carriers of heat. For each concentration, $n$, a temperature, $T_{qn} = hv_{s}/\ell_{dd}k_{B}$, can be defined, which corresponds to equality between the typical acoustic phonon wavelength, $\lambda_{ph}=2\pi/q_{ph}$  and interdopant distance, $\ell_{dd}=n^{-1/3}$.  As one can see in Fig. 3c,  $T_{qn}$ is close to where $\Delta\kappa_{ph}$ becomes large. Such a crude picture based on the Debye approximation, should not be taken too literally  in presence of soft modes.

In principle, \emph{Ab Initio} calculations\cite{Lindsay:2013} can give an account of heat transport near room temperature. Recently,  two groups \cite{Feng:2015,Tadano:2015} succeeded in determining the phonon spectrum of strontium titanate free of the commonly-found imaginary frequencies\cite{Aschauer:2014} and computing the intrinsic lattice conductivity of the cubic phase. Fig. 4a compares our high-temperature data with these calculations \cite{Feng:2015,Tadano:2015} as well as previous experimental reports\cite{Steigmeier:1968,Muta:2005,Yu:2008}. As one can see in the figure, there is a broad agreement between experimental results. Theoretical calculation using the Generalized Gradient Approximation (GGA)\cite{Feng:2015}  are very close to the experimental data above 250 K. On the other hand, the experimental slope matches more the theory based on microscopic anharmonic force constants\cite{Tadano:2015}.

\begin{figure}
\includegraphics[width=0.4\textwidth]{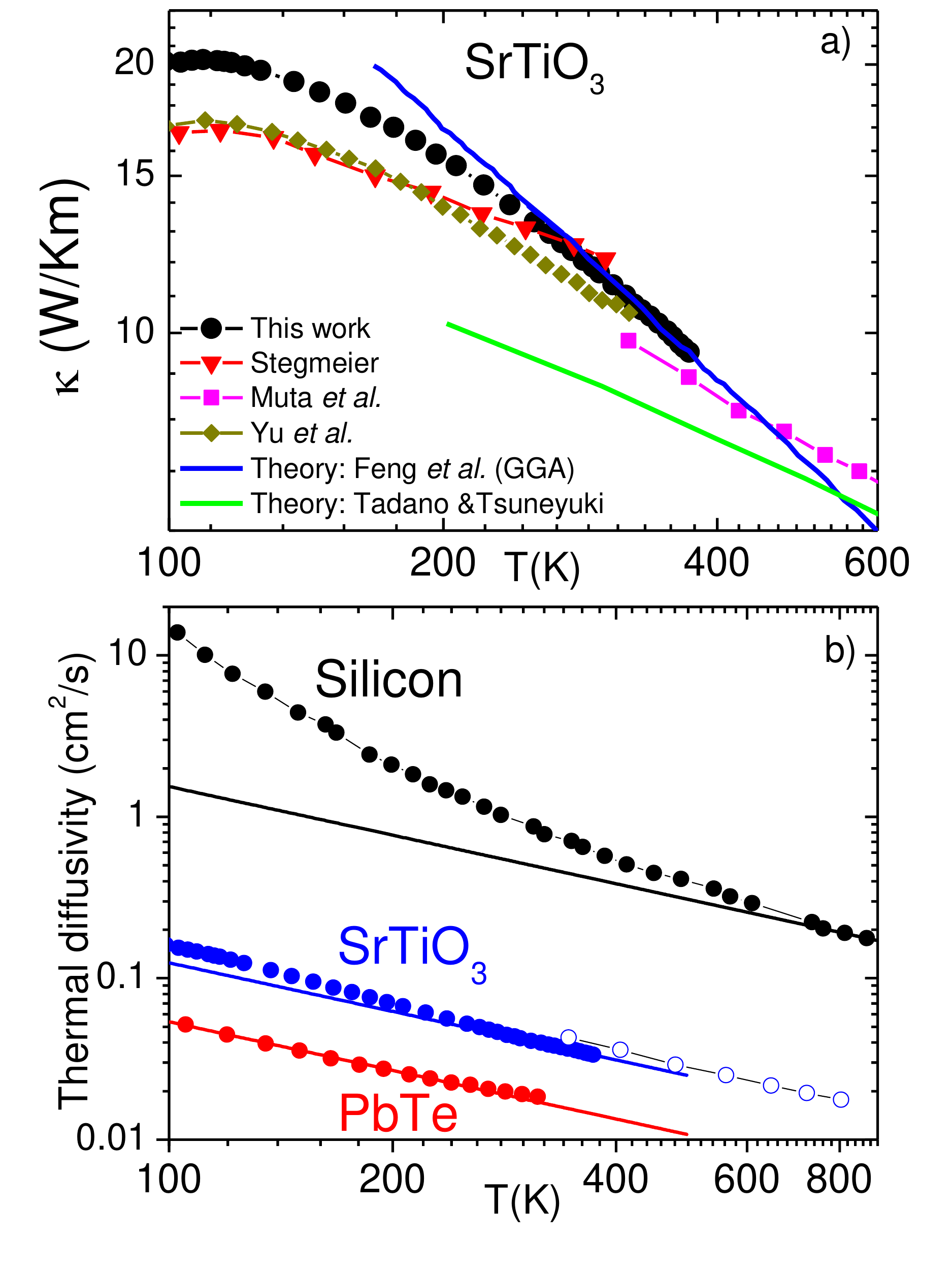}
\caption{ a) Thermal conductivity at high-temperature compared to previous experimental reports \cite{Steigmeier:1968,Muta:2005,Yu:2008} and  theoretical calculations\cite{Feng:2015,Tadano:2015}; b) Thermal diffusivity, $D$, extracted from thermal conductivity and  specific heat data as a function of temperature in SrTiO$_{3}$ (solid blue circles) together with data from ref.\cite{Hofmeister:2010} (open circles), compared to silicon and PbTe. Solid lines represent $D= s v_{s}^{2}\tau_{P}$ (See text).}
\label{Fig4}
\end{figure}

Let us conclude by a short discussion of thermal diffusivity, $D =\frac{\kappa}{C_{p}}$ in this regime. We can extract $D$  by combining our thermal conductivity data and the specific heat. Fig. 4b presents the temperature dependence of thermal diffusivity. One can see that, at room temperature and above, thermal diffusivity tends to be proportional to $T^{-1}$. Our data is in good agreement with reported values of thermal diffusivity at high temperature\cite{Hofmeister:2010}. In the vicinity of room temperature and above, thermal diffusivity becomes proportional to the inverse of temperature. The thermal diffusivity of a good conductor of heat, silicon and a very bad one,  PbTe, are also shown. Remarkably, in the two bad conductors, the magnitude and the temperature dependence of $D$ in the high-temperature regime can be expressed in a very simple way:

\begin{equation}
D=  s v^2_{s} \tau_{p}
\end{equation}

Here, $\tau_{p}=\frac{\hbar}{k_{B}T}$ is the Planckian scattering time\cite{Bruin:2013}, and $s$ a dimensionless parameter (See table I). In PbTe and SrTiO$_{3}$, $s$ is close to unity and the temperature dependence is set by $\tau_{p}$.  It emerges as a useful parameter for comparing the thermal conductivity of different cubic insulators. In many perovskytes, recently studied by Hofmeister\cite{Hofmeister:2010}, $D$ has a comparable magnitude and temperature dependence. On the other hand, in a highly conducting cubic insulator such as silicon, $D$ is much larger and drops faster, presumably because  the phase space for three-phonon scattering Umklapp events\cite{Lindsay:2008} is smaller.

Eq. 1 is strikingly similar to the suggested universal boundary on diffusivity suggested by Hartnoll\cite{Hartnoll:2015}, with sound velocity replacing the Fermi velocity.  The experimental motivation for Hartnoll's proposal\cite{Hartnoll:2015} was the fact that $\tau_{p}$ is the average scattering rate of electrons in numerous metals with linear resistivity\cite{Bruin:2013}. Is there  a boundary to thermal transport by phonons in insulators? In other, words, is there a fundamental reason for $s$ to remain larger than unity? These are the questions raised by our observation.

JLJ acknowledges the Science Without Borders program of CNPq/MCTI-Brazil and VM and KB acknowledge FAPERJ fellowships (Nota 10 and Visitante). KB is also supported by Fonds ESPCI and by a QuantEmX grant (GBMF5305) from ICAM and the Gordon and Betty Moore Foundation. We thank B. Fauqu\'e, Y. Fuseya, S. A. Hartnoll and A. Kapitulnik for stimulating discussions.

\begin{table}[htbp]\centering
\label{tab:2}
\renewcommand\arraystretch{1.6}
\renewcommand\tabcolsep{4pt}
\begin{tabular}{| c | c | c |c | }\hline
system   & D$_{300K}$ (mm$^{2}$/s) & v$_{sl}$(100)(km/s)& s  \\
\hline
SrTiO$_{3}$ &4.0& 7.87& 2.6\\
\hline
PbTe & 1.9 & 3.59 & 5.9\\
\hline
Si & 91 & 8.43 &  51 \\
\hline
\end{tabular}
\caption{Room-temperature thermal diffusivity ($D_{300K}$), longitudinal sound velocity ($v_{sl}$)  and the parameter $s$  quantifying the slope of high-temperature thermal diffusivity, in three cubic solids. }
\end{table}

\newpage
\appendix
\section{Methods}
Thermal conductivity and specific heat measurements were performed into the commercial system Dynacool PPMS (Physical Property Measurement System) that allows to perform experiments between 1.7 and 400K.

Thermal conductivity was measured with a standard two-thermometers-one heater configuration. The power supplied through the heater established a temperature difference
 ($\Delta T=T_1-T_2$) along the sample that was kept below 1$\%$ of the average temperature ($T_{av}=(T_1+T_2)/2$). This condition guarantees to have a negligible thermal flow along the electrical wires of the sensors and the heater, and to obtain an accurate determination of the thermal conductivity. The major source of experimental error comes from the determination of the geometrical factor that can be up to 6$\%$.

Specific heat was measured with the standard commercial platform compatible with Dynacool. Temperature rises were limited to 1-2\% of base temperature. Due to the low heat capacity of SrTiO$_{3}$ at the lowest temperature, a minimum sample mass of 45mg was necessary to obtain a sizeable contribution of the sample heat capacity respect to the addenda (N-apiezon).

\section{Samples} The samples investigated (SrTiO$_{3}$, SrTi$_{1-x}$Nb$_{x}$O$_{3}$, Sr$_{1-x}$Ca$_{x}$TiO$_{3}$ and KTaO$_{3}$) in this work are all commercial single crystal specimens. For SrTiO$_{3}$, two of the three samples of SrTiO$_{3}$ came from two different batches of the same  supplier (sample-1 and sample-3), whereas the other one came from a second supplier (sample-2).

The carrier density was determined by measuring the Hall resistivity and was found to be in good agreement with the nominal Nb content. Table I summarizes  relevant values of both electrical and thermal transport measurements. $\rho_0$ and $A$ are obtained fitting the resistivity curve at the lowest measured temperature with the function $\rho=\rho_0$+AT$^2$ (see solid lines in Fig. \ref{Fig:R}b).

\begin{table*}[htbp]\centering
\label{tab:1}
\renewcommand\arraystretch{1.6}
\renewcommand\tabcolsep{4pt}
\begin{tabular}{| c | c | c | c | c | c | c | c | c | c | }\hline
\textbf{x}  & \textbf{$n$} & \textbf{$\rho_{300K}$} & \textbf{$\rho_{2K}$} & RRR & \textbf{$R_H$} & \textbf{$\rho_0$} & \textbf{$A$} & \textbf{$\kappa_{300K}$} & \textbf{$\kappa_{2K}$} \\
\textbf{}  & (cm$^{-3}$) & (m$\Omega$cm) & (m$\Omega$cm) & \textbf{} & (cm$^{-3}$/C) & (m$\Omega$cm) & ($\mu$$\Omega$cm/K$^2$) & (W/cmK)  & (W/cmK)\\ \hline \hline
0.0004 & 0.53(4)$\cdot 10^{19}$ & 277 & 0.088 & 2579 &1.18& 0.078(4)&0.85(1)&0.1&0.0013\\
0.001 & 1.4(1)$\cdot 10^{19}$ & 59.1 & 0.08&  739& $4.33\cdot 10^{-1}$& 0.077(4)&0.26(1)&0.099&0.0011\\
0.01 & 9.4(7)$\cdot 10^{19}$ & 9.46 & 0.064 & 148 &$6.59\cdot 10^{-2}$& 0.060(2)&0.062(1)&0.093&7.5$\cdot10^{-4}$\\
0.014 & n.a. & 4.22 & 0.087 & 49 & n.a. &0.075(5) & 0.030(8) &0.098&0.0014\\
0.02 & 2.5(9)$\cdot 10^{20}$ & 3.53 & 0.077 & 46& $2.45\cdot 10^{-2}$ &0.072(4) & 0.029(8)&0.088&0.0011\\
\hline\end{tabular}
\caption{The samples SrTi$_{1-x}$Nb$_x$O$_3$ investigated in the present work. The table reports on the nominal content of Nb per formula unit ($x$), the charge carrier concentration ($n$) determined by Hall resistivity, resistivity at 300K ($\rho_{300K}$), resistivity at 2K ($\rho_{2K}$), RRR = $\frac{\rho_{300K}}{\rho_{2K}}$, Hall coefficient ($R_H$), $\rho_0$ and $A$ coefficients obtained fitting the low temperature data with the function $\rho=\rho_0+A\cdot T^2$, thermal conductivity at 300K ($\kappa_{300K}$), thermal conductivity at 2K ($\kappa_{2K}$).}
\end{table*}

\section{Thickness dependence of thermal conductivity}
\begin{figure*}
	\includegraphics[width=0.8\textwidth]{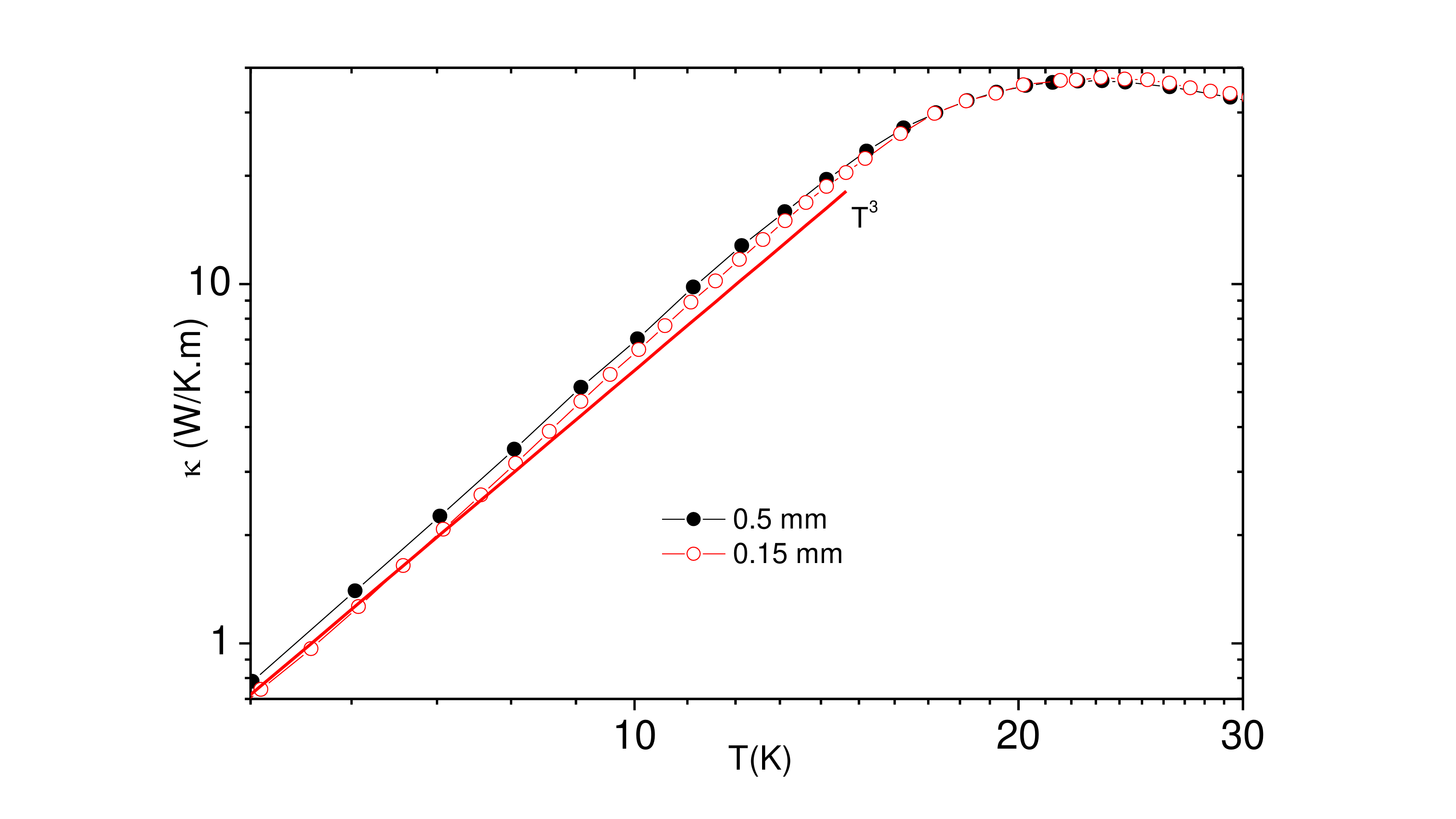}
	\caption{Thickness dependence thermal conductivity of SrTiO$_3$: sample thickness $t=$0.5mm (full black circles) and $t=$0.15mm (open red circles). The two curves differ only in the Poiseuille regime.}
	\label{Figs5:size}
\end{figure*}

Sample-3, initially 500$\mu$m-thick, was thinned down to 150$\mu$m in order to perform the thickness-dependence measurement of thermal conductivity (Fig. \ref{Figs5:size}). We observed that the thermal conductivity decreases in the ballistic regimes and in the temperature range that we identified as Poiseuille, but not at the peak temperature and above.

The lower thermal conductivity at low temperatures implies a lower mean free path, when phonon thermal conductivity is expressed as $k=\frac{1}{3}c_{ph}v_{ls}$. As discussed in the main text, the mean free path of SrTiO$_3$ in the ballistic regime is much lower than the sample size and of the order of magnitude of the domain size[S1]. Therefore, the modest reduction of the mean free path by reducing thickness suggests a either a correlation between domain and sample dimensions or the existence of a small subset of phonons which can travel across domain boundaries.

\section{Thermal conductivity and its sensitivity to disorder}
\begin{figure*}
\includegraphics[width=0.8\textwidth]{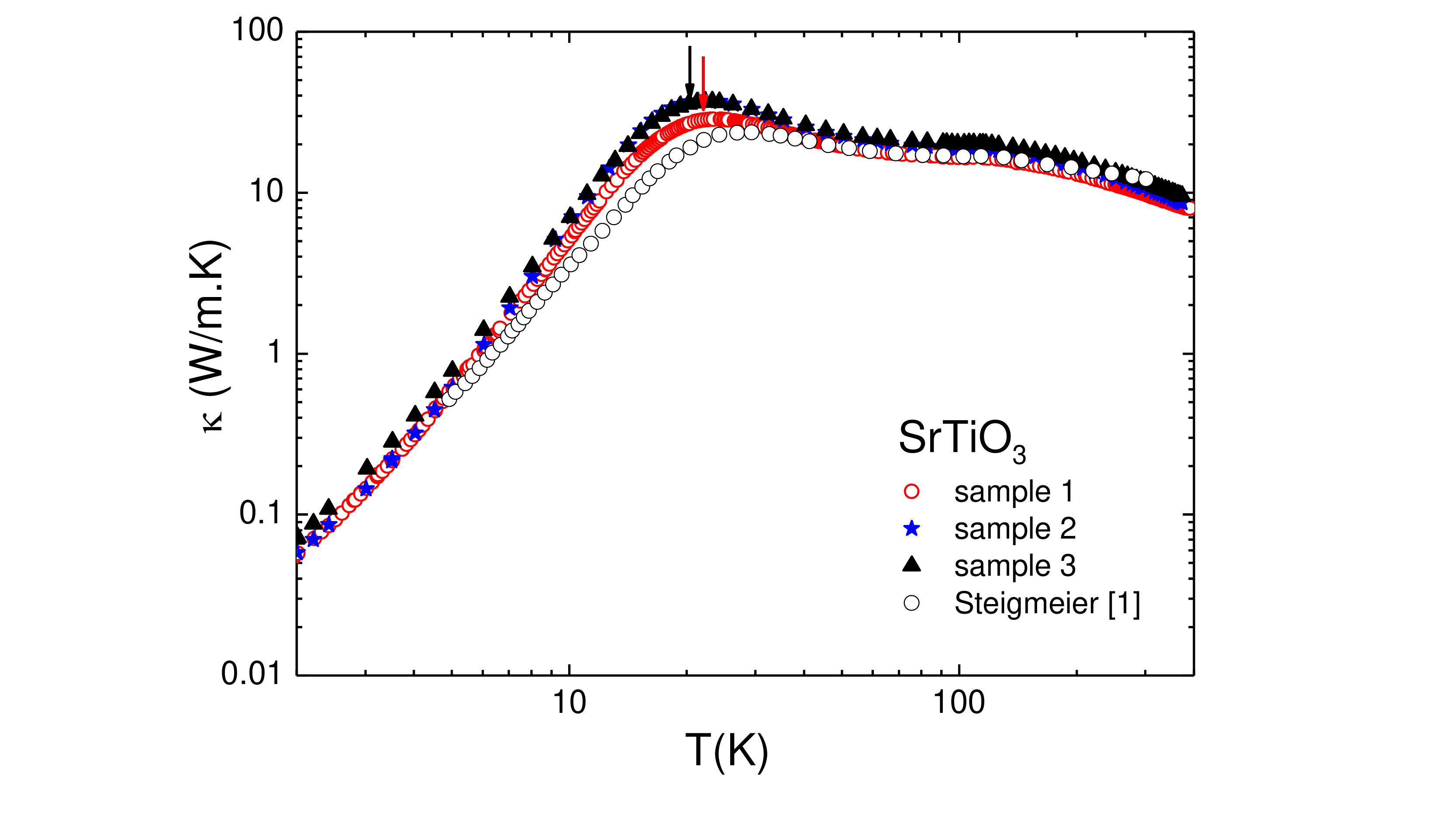}
\caption{Thermal conductivity in three SrTiO$_3$ samples compared with the data of reference \citep{Steigmeier:1968}. Sample 1 and 3 come from the same provider. The downward arrows are placed at temperatures where thermal conductivity shows a maximum ($\kappa_{peak}$).}
\label{Figs1:k}
\end{figure*}

Fig. \ref{Figs1:k} presents the thermal conductivity of the three samples compared with an early study [S2].  One can see an overall agreement between the data sets. The peak thermal conductivity, $\kappa_{peak}$ (showed by downward arrows), appears to be sample-dependent. Since controlled substitution drastically affects $\kappa_{peak}$ (see below), it is reasonable to assume that $\kappa_{peak}$ is very sensitive to disorder and is largest in the cleanest samples.

\begin{figure*}
\includegraphics[width=0.8\textwidth]{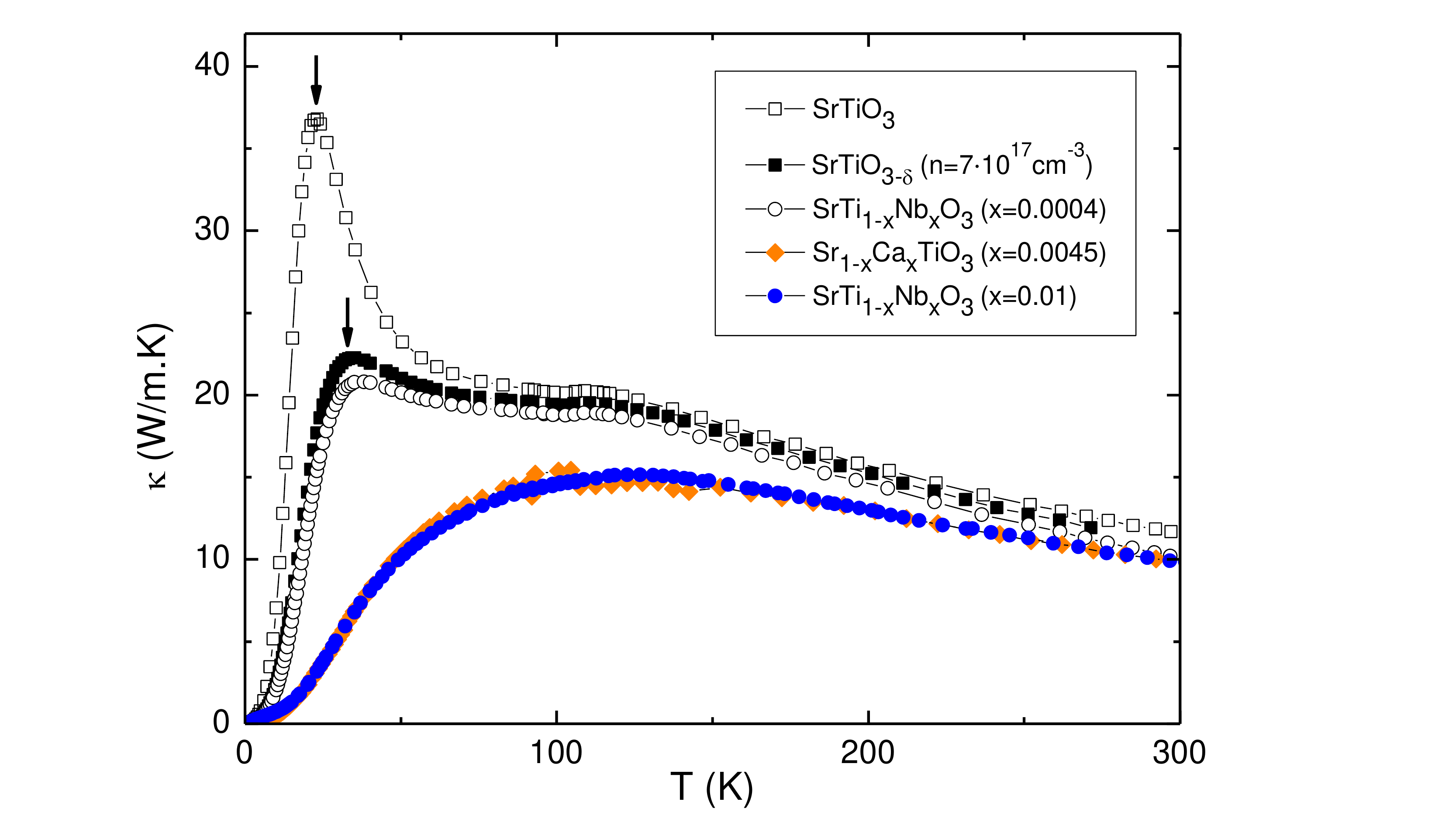}
\caption{Thermal conductivity of  SrTiO$_3$ (sample 3) compared with a number of samples with different atomic substitutions. Note the strong damping of the amplitude of $\kappa_{peak}$ (represented by downward arrows) with the introduction of a tiny amount of extrinsic atoms.}
\label{Fig:k_s}
\end{figure*}

Fig. \ref{Fig:k_s} compares the thermal conductivity of one of our cleanest SrTiO$_3$ samples with a reduced (SrTiO$_{3-\delta}$ ), two niobium-doped and one calcium substituted sample. One can see that the effect of oxygen reduction and Ca substitution is similar to the effect of Nb substitution, which was studied \emph{in extenso} (See Fig. 3 of the main text). An extremely low level of atomic substitution drastically reduces the $\kappa_{peak}$. Ca substitution, oxygen reduction and Nb substitution have a very similar conseqiences. Now, the latter two  turn the system to a metal, but the former does not introduce mobile electrons). Therefore, as argued in the main text, the main reason for the observed reduction in lattice thermal conductivity is not electron scattering but disorder.

\begin{figure*}
\includegraphics[width=0.6\textwidth]{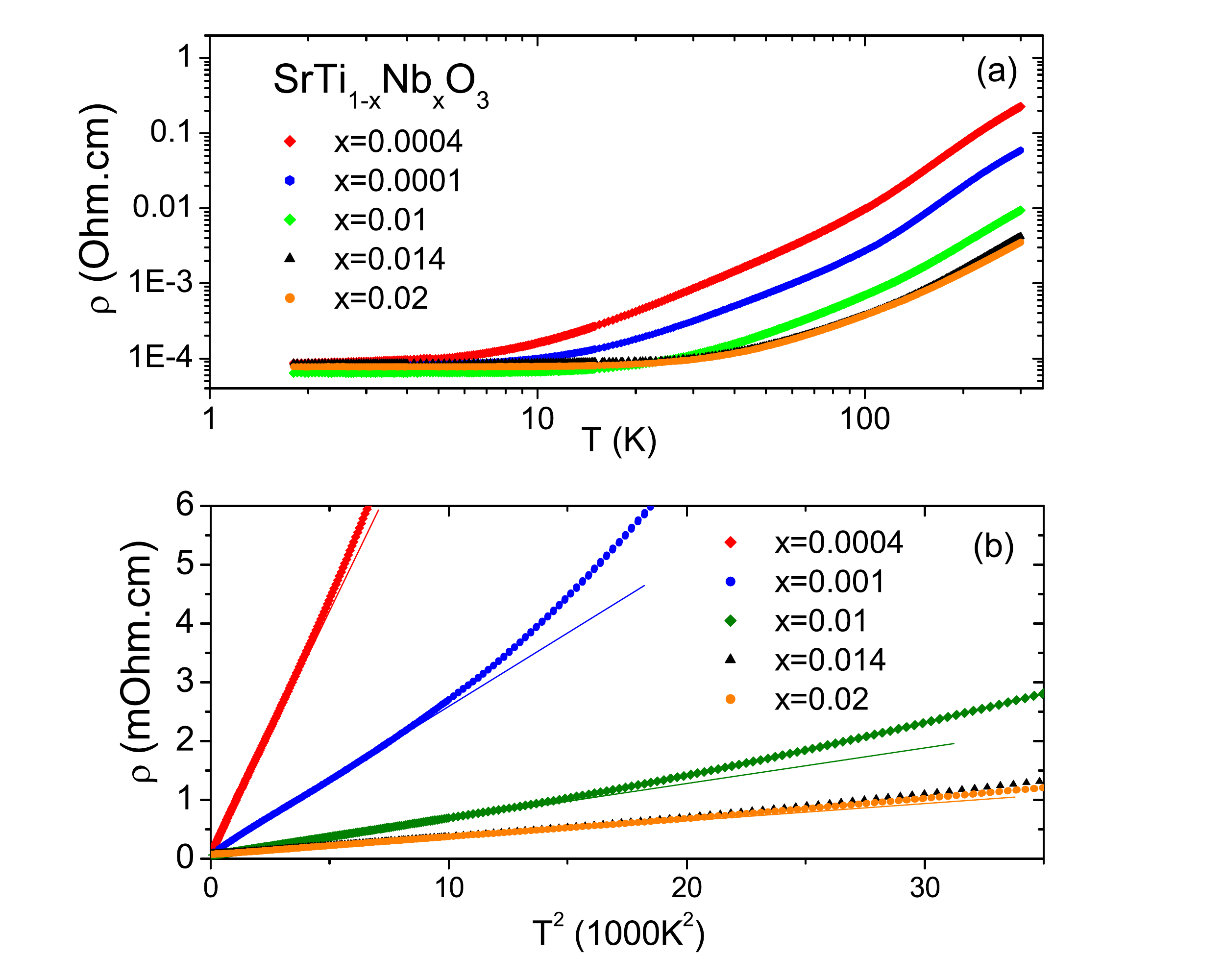}
\caption{Panel(a): Electrical resistivity between 1.7 and 300K as a function of temperature for four different samples. Panel (b): low temperature electrical resistivity as a function of $T^2$, the solid line is a guide for the eye representing the  T$^2$ behaviour at low temperature. Above a finite temperature, resistivity deviates upward from this quadratic behavior.  quadratic behaviour widens with increasing doping [S4].}
\label{Fig:R}
\end{figure*}

\section{Electric resistivity} Electrical resistivity and Hall-resistivity were also carried out in the PPMS system. The data are similar to what was reported previously [S3, S4]. Figure \ref{Fig:R} shows the resistivity of SrTi$_{1-x}$Nb$_{x}$O$_{3}$ as a function of temperature (panel (a)) and as a function of $T^2$ (panel (b)). As found previously[S4], resistivity follows a $T^2$ behavior at low temperature and then a faster than $T^2$ (close to cubic) at higher temperatures.

\begin{figure*}
\includegraphics[width=0.8\textwidth]{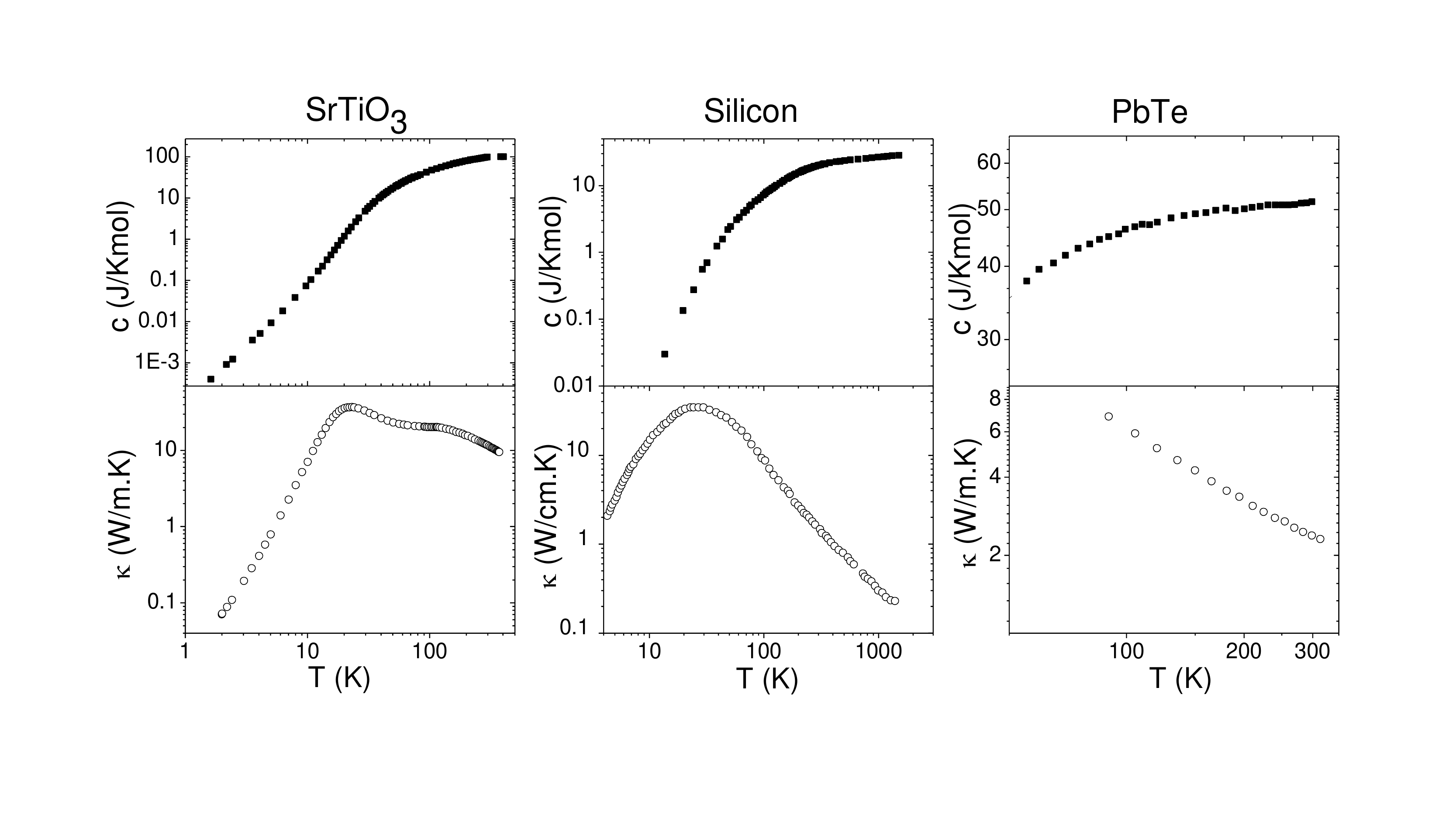}
\caption{Specific heat and thermal conductivity for strontium titanate, silicon and lead telluride. For silicon, the specific heat was reported in Ref. S5 and thermal conductivity in Ref. S6. For PbTe, the specific heat was reported in S7 and S8 and thermal conductivity in  S9.}
\label{Fig:D}
\end{figure*}

\section{Thermal diffusivity in three cubic solids}
Figure \ref{Fig:D} shows the thermal conductivity and specific heat of SrTiO$_{3}$, Si and PbTe.  This data was used to extract the thermal diffusivity of these three cubic semiconductors shown in Fig. 4  of the main text. To quantify the specific heat per volume, the molar volume was used as specified in Table II. The magnitude of $s$ is the main text was extracted using equation 1 and taking v$_s$ to be the longitudinal sound velocity along 100 as specified in the table.

\begin{table*}[htbp]\centering
\label{tab:2}
\renewcommand\arraystretch{1.6}
\renewcommand\tabcolsep{4pt}
\begin{tabular}{| c | c | c | c | c | c | c | }\hline
system   & V$_m$(cm$^{3}$/mol & v$_{sl}$(100) (km/s)&  v$_{st}$(100) (km/s) & $\kappa_{300K}$(W/m.K)& C$_{300K}$ (J/cm$^{3}$K) & D$_{300K}$ (cm$^{2}$/s) \\
\hline
SrTiO$_{3}$  & 35.7 & 7.87 & 4.9 & 11.0 & 2.75 & 0.04 \\
\hline
Si & 12.1 & 8.43 &  5.84 & 150 & 1.65 & 0.91 \\
\hline
PbTe & 40.9 & 3.59 &  1.26 & 2.37 &1.25 & 0.019 \\
\hline
\end{tabular}
\caption{Molar volume ($V_m$), longitudinal ($v_{sl}$) and transverse sound velocity ($v_{st}$) in three cubic solids together with their room-temperature thermal conductivity ($\kappa_{300K}$), specific heat ($c_{300K}$) and thermal diffusivity ($D_{300K}$).}
\end{table*}

\section{References}
\textbf{S1:} A. Buckley, J. P. Rivera and E. K. H. Salje, J. Appl. Phys.  \textbf{86}, 1653 (1999).

\textbf{S2}: E. F. Steigmeier,  Phys. Rev. \textbf{168}, 523 (1968).

\textbf{S3:} A. Spinelli,  M. A. Torija, C. Liu, C. Jan and C. Leighton, Phys. Rev. B \textbf{81}, 155110 (2010).

\textbf{S4:} X. Lin, B. Fauqu\'e and K. Behnia, Science \textbf{349}, 945 (2015).



\textbf{S5:} P. D. Desai, J Phys. Chem Ref. Data \textbf{15}, 967(1985).

\textbf{S6:} C. J. Glassbrenner and Glen, Phys. Rev. \textbf{134}, A1058, (1964).

\textbf{S7:} D. H. Parkinson and J. E. Quarrington. Proc. Phys. Soc. 67, 569 (1954)

\textbf{S8: }A. S. Pashinkin \emph{et al.}, Inorganic Materials \textbf{45}, 1226 (2009).

\textbf{S9:} D. T. Morelli, V. Jovovic and J. P. Heremans, Phys. Rev. Lett. \textbf{101}, 035901 (2008).

\end{document}